\documentclass[a4paper]{jpconf}

\usepackage{citesort}
\usepackage{subfigure}
\usepackage{footnote}
\usepackage{url}
\usepackage{graphicx}
\usepackage{url}
\begin{document}
\title{Latest MAGIC discoveries pushing redshift boundaries in VHE Astrophysics}

\author{M Manganaro$^1$, J Becerra$^2$, M Nievas$^3$, J Sitarek$^4$, F Tavecchio$^5$, S Buson$^6$, D Dominis$^7$, A Dom\'inguez$^3$, E Lindfors$^8$, D Mazin$^9$, A Moralejo$^{10}$, A Stamerra$^5$ and Ie Vovk$^{11}$ for the MAGIC and FERMI collaboration}

\address{$^1$ Inst. de Astrof\'isica de Canarias, E-38200 La Laguna, and Universidad de La Laguna, Dpto. Astrof\'isica, E-38206 La Laguna, Tenerife, Spain}
\address{$^2$ NASA Goddard Space Flight Center, Greenbelt, MD~20771, USA and Department of Physics and Department of Astronomy, University of Maryland, College Park, MD~20742, US}
\address{$^3$ Universidad Complutense, E-28040 Madrid, Spain}
\address{$^4$ University of Lodz, PL-90236 Lodz, Poland}
\address{$^5$ INAF National Institute for Astrophysics, I-00136 Rome, Italy}
\address{$^6$ Universit\`a di Padova and INFN, I-35131 Padova, Italy}
\address{$^7$ Croatian MAGIC Consortium, Rudjer Boskovic Institute, University of Rijeka and University of Split, HR-10000 Zagreb, Croatia}
\address{$^8$ Finnish MAGIC Consortium, Tuorla Observatory, University of Turku and Department of Physics, University of Oulu, Finland}
\address{$^9$Japanese MAGIC Consortium, ICRR, The University of Tokyo, Department of Physics and Hakubi Center, Kyoto University, Tokai University, The University of Tokushima, KEK, Japan}
\address{$^{10}$ IFAE, Campus UAB, E-08193 Bellaterra, Spain}
\address{$^{11}$ Max-Planck-Institut fur Physik, D-80805 Munchen, Germany}

\ead{manganaro@iac.es}

\begin{abstract}

The search for detection of $\gamma$-rays from distant AGNs by Imaging Atmospheric Cherenkov Telescopes (IACTs) is challenging at high redshifts, not only because of lower flux due to the distance of the source, but also due to the consequent absorption of $\gamma$-rays by the extragalactic background light (EBL). Before the MAGIC discoveries reported in this work, the farthest source ever detected in the VHE domain was the blazar PKS~1424+240, at $z>0.6$. MAGIC, a system of two 17 m of diameter IACTs located in the Canary island of La Palma, has been able to go beyond that limit and push the boundaries for VHE detection to redshifts $z\sim 1$. The two sources detected and analyzed, the blazar QSO~B0218+357  and the FSRQ PKS~1441+25  are located at redshift $z=0.944$ and $z=0.939$ respectively. QSO~B0218+357 is also the first gravitational lensed blazar ever detected in VHE. The activity, triggered by {\it Fermi}-LAT in high energy $\gamma$-rays, was followed up by other instruments, such as the KVA telescope in the optical band and the {\it Swift}-XRT in X-rays. In the present work we show results on MAGIC analysis on QSO~B0218+357 and PKS~1441+25 together with multiwavelength lightcurves. The collected dataset allowed us to test for the first time the present generation of EBL models at such distances. 

\end{abstract}

\section{Introduction}

The Very High Energy (VHE, $>$100GeV) sky is populated by almost 60 blazars, active galactic nuclei which relativistic jet is pointing in the direction of the Earth. The peculiar orientation of the jet towards the observer, and their features such as continuum emission, rapid variability, high polarization, and radio structures dominated by compact radio cores, make them a perfect laboratory to study the nature of the emission at different wavelengths, from radio up to the TeV range, and to investigate, at the highest energies, the effect of the EBL absorption on the $\gamma$-rays produced by the central engine of the galaxy. Until recently, the farthest blazar detected at VHEs was PKS~1424+240, at $z>0.6$ \cite{Acciari_2010}.
Thanks to the recent discoveries in VHE by MAGIC of QSO~B0218+357 \cite{ATel6349} and PKS~1441+25 \cite{ATel7416}, at respectively $z=0.944$ \cite{Linford_2012} and $z=0.939$ \footnote{From SDSS: \url{http://skyserver.sdss.org/dr10/en/get/SpecById.ashx?id=6780257851631206400} and \cite{shaw12}} the VHE sky suddenly expanded to its limits reaching the unprecedented redshift $z\sim 1$. Beyond the so-called cosmic $\gamma$-ray horizon, the Universe becomes opaque to VHE $\gamma$-ray radiation due to the interaction of high energy $\gamma$-rays with the diffuse EBL, producing electron-positron pairs. At the redshift of $z\sim 1$ this absorption results in a cut-off at the energy of $\sim$100 GeV \cite{Dominguez13,Blanch2005}. Such energies are at the lower edge of the performance of the current generation of Imaging Atmospheric Cherenkov Telescopes (IACTs), and only a low threshold of the instrument can make the detection and the consequent study of the source possible. To maximize the detection chance, the observations are often triggered by a high state observed in lower energy ranges. The Large Area Telescope on board {\it Fermi}, scanning the whole sky in the High Energy (HE, $0.1 \,\rm{GeV}<E<100$\,GeV) range provides alerts of high energy fluxes and spectral shape triggering MAGIC observations, as for the two sources we are describing in this work.

\section{Instruments in brief}
MAGIC is a stereoscopic system consisting of two 17\,m diameter Imaging Atmospheric Cherenkov Telescopes located at the Observatorio del Roque de los Muchachos, on the Canary Island of La Palma.
The current integral sensitivity for low-zenith observations ($zd<30^\circ$) above 100\,GeV is $1.445\pm0.015\,\%$ of the Crab Nebula's flux in 50~h, while the one above 220\,GeV is $0.66\pm0.03\,\%$. The Field of View (FoV) of MAGIC is $3.5^\circ$ \cite{Aleksic2015_Upgrade2}.
The Large Area Telescope (LAT) on board of {\it Fermi} is an imaging high-energy $\gamma$-ray telescope in the energy range from about 20 MeV to more than 300 GeV \cite{Atwood2009}. The LAT's FoV covers about 20\% of the sky at any time, and it scans continuously, covering the whole sky every three hours. 
For both sources, multiwavelength(MWL) observations were performed, collecting data from other instruments.
{\it Swift}-XRT \cite{Burrows2005} in the X-ray and KVA\footnote{\url{http://users.utu.fi/kani/1m}} in the R-band were collected for QSO~B0218+357 and PKS~1441+25  while for PKS~1441+25 only, data from NuSTAR \cite{Harrison13} in Hard X-ray, {\it Swift}-UVOT \cite{Roming2005} in optical-UV, Hans-Haffner\footnote{\url{http://schuelerlabor-wuerzburg.de/?p=Sternwarte}} in the optical R-band, CANICA\footnote{\url{http://www.inaoep.mx/~astrofi/cananea/canica/}} in the Near Infrared and Mets\"{a}hovi for Radio were used \cite{Terasranta}. 

\section{Observations and Results}

\subsection{QSO~B0218+357}
\begin{figure}[h]
\centering
\subfigure[$\theta^{2}$ for QSO~B0218+357]{%
\includegraphics[width=16pc]{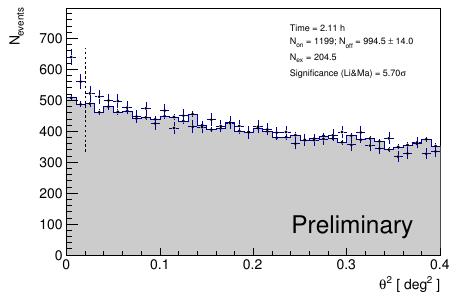}
\label{fig:th2_1441}}
\quad
\subfigure[$\theta^{2}$ for PKS~1441+25]{%
\includegraphics[width=16pc]{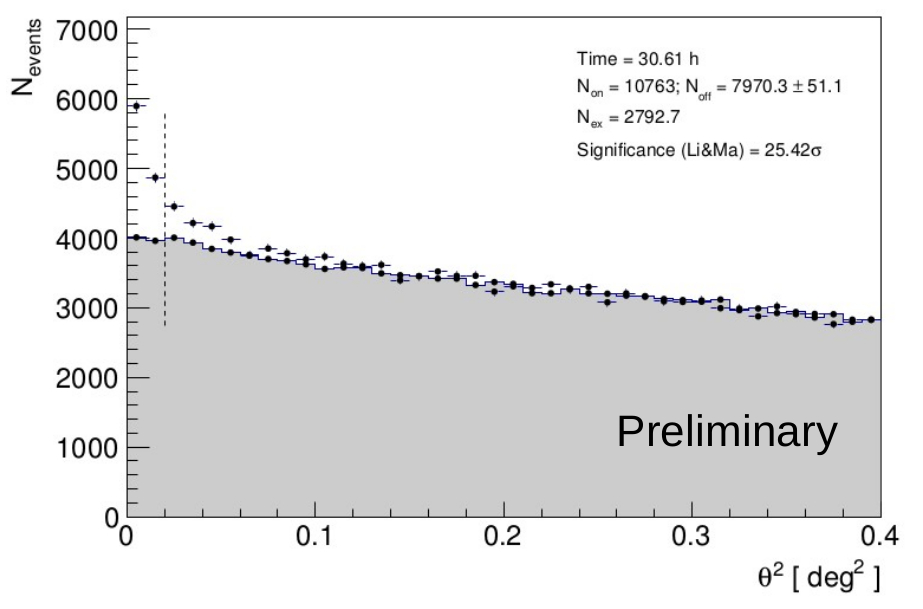}
\label{fig:th2_QSO}}
\caption{Distribution of the squared angular distance $\theta^{2}$ between the reconstructed source position and the nominal source position (points) or the background estimation position (shaded area) for QSO~B0218+357 (a) and  PKS~1441+25 (b). Vertical dashed line shows the value of $\theta^{2}$ up to which the number of excess events and significance are computed.}
\label{fig:th2}

\end{figure}

QSO~B0218+357 is a blazar located at the redshift of 0.944 \cite{Linford_2012}. The object is gravitationally lensed by a galaxy [PBK93]~B0218+357 G located at $z=0.68$. The time delay between the two flares, measured by {\it Fermi}-LAT during a previous flare in 2012, was expected to be $11.46\pm 0.16$ days. 
In July 2014 a flare observed by {\it Fermi}-LAT \cite{Cheung2014} triggered follow-up observations of the MAGIC telescopes which in turn led to the discovery of VHE $\gamma$-ray emission from QSO~B0218+357 \cite{ATel6349} from the secondary component of the flare, observed at the expected time of arrival with a significance of 5.7$\sigma$ (see Fig.\ref{fig:th2}, panel a). As in the case of PKS~1441+25 the analysis of the data, performed within the standard MAGIC analysis framework \emph{MARS} \cite{Zanin2013_MARS,Aleksic2015_Upgrade2} was carried on with the aim of ad-hoc Monte Carlo (MC) simulations to match the night-sky background.
The observations were performed in 14 nights in July and August 2014 for a total duration of 12.8~h at intermediate zenith angle ($20^\circ<zd<43^\circ$). 

The preliminary MWL lightcurve of the source is shown in Fig.\ref{MWL_QSO}. The light curve obtained by MAGIC above 100 GeV is compared with the emission observed above 0.3 GeV by {\it Fermi}-LAT, {\it Swift}-XRT in 0.3-10 keV range and in the bottom panel by KVA in the R-band. The {\it Fermi}-LAT emission during the second component of the flare was a factor 20 higher than the average state of the source \cite{Acero15}. During both components of the flare the photon index was significantly harder \cite{Buson14} than the average one. The MAGIC emission occurred at the same time as the delayed, weaker component of the flare in {\it Fermi}-LAT. The two flaring nights give the mean flux of $(5.8\pm 1.6_{stat} \pm 2.4_{syst} ) \times 10^{-11} \rm{cm}^{-2} s^{-1}$ above 100 GeV \cite{Sitarek15}. 

The spectral features of the emission, modelling and impact on the EBL measurements will be discussed in \cite{Ahnenprep}.

\subsection{PKS~1441+25}
The MAGIC telescopes monitored PKS~1441+25 in April and May 2015 for a total of $\sim30~$h. The observations were performed in the standard wobble mode with a $0.4^{\circ}$ offset and four symmetric positions with respect to the camera center. The data were collected in the zenith angle range of $3^\circ<zd<38^\circ$.
The analysis of the data was performed using the standard MAGIC analysis framework \emph{MARS} \cite{Zanin2013_MARS,Aleksic2015_Upgrade2} and, as in the case of QSO~B0218+357, with Monte Carlo simulations matching the night-sky background levels. 
PKS~1441+25 was detected with a significance of $25.5\,\sigma$, see Fig.\ref{fig:th2}, panel b.

\begin{figure}[h]
\begin{minipage}{17pc}
\includegraphics[width=17pc]{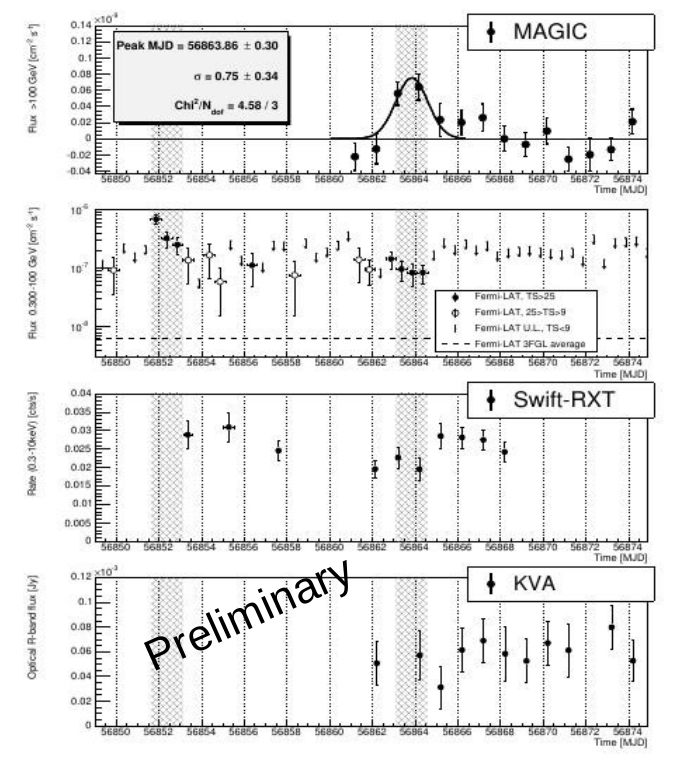}
\caption{\label{MWL_QSO}MWL for QSO~B0218+357, during the flaring state in July/August 2014. Top panel: MAGIC (points) above 100 GeV and Gaussian fit to the peak position (thick solid line). Second panel from the top: {\it Fermi}-LAT above 0.3 GeV (note the log scale) with the average flux from the 3rd {\it Fermi} Catalog \cite{Acero15} marked with dashed line. Third panel from the top: {\it Swift}-XRT in 0.3-10 keV range from the automatic analysis \cite{Stroh13}. Bottom panel: KVA in R band.}
\end{minipage}\hspace{2pc}%
\begin{minipage}{20pc}
\includegraphics[width=20pc]{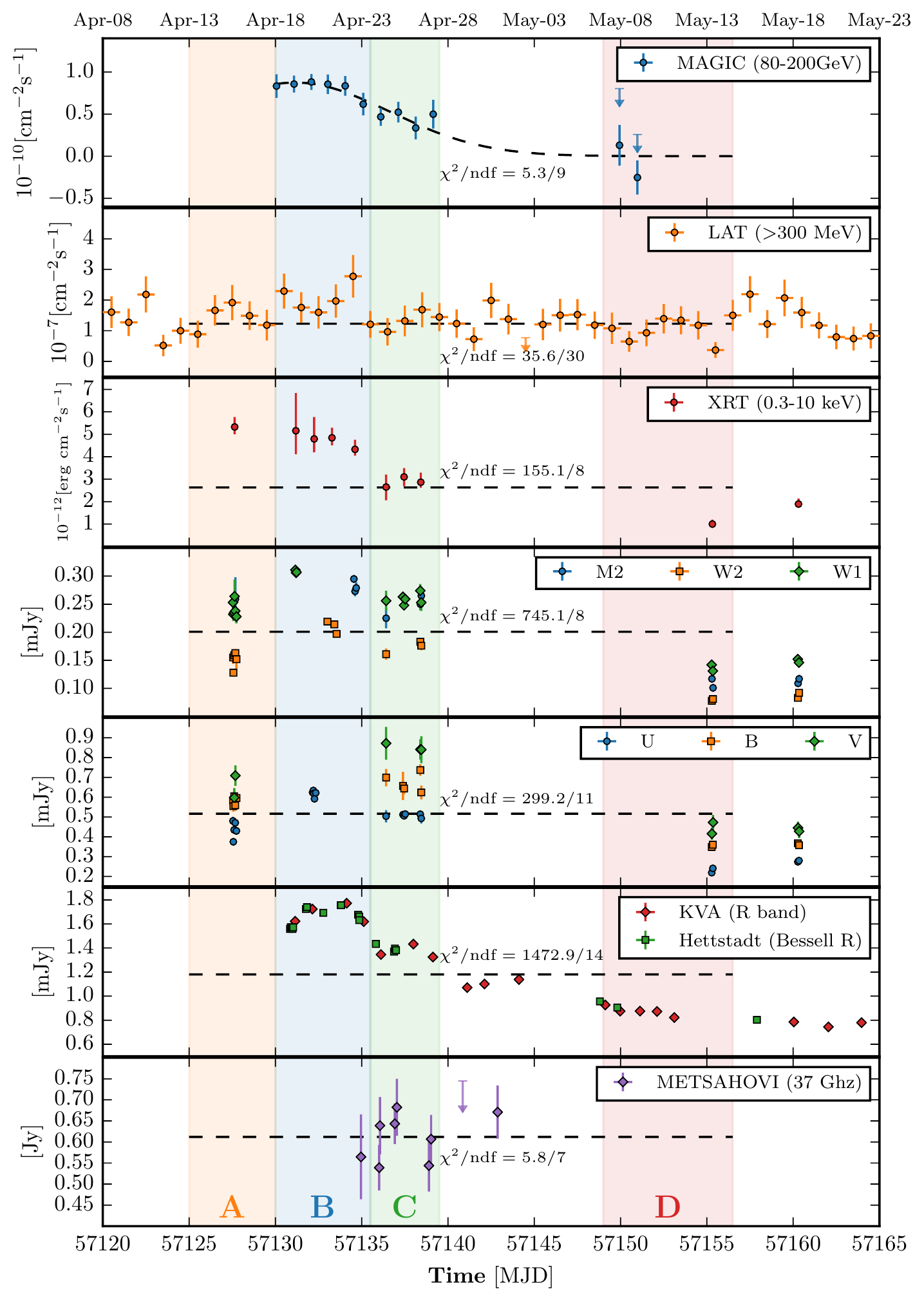}
\caption{\label{MWL_1441}MWL lightcurve for PKS~1441+25. The shaded areas marked as A, B, C, D depict four different states of the source as in \cite{Ahnen15}.}
\end{minipage} 
\end{figure}

The MWL lightcurve for PKS~1441+25 is shown in Fig.\ref{MWL_1441}. The differential VHE spectrum was measured from 40 to 250\,GeV. Employing state-of-the-art EBL models, upper limits to the EBL density are derived and reported in \cite{Ahnen15}: the upper limits on the optical depth ($\tau$) calculated under the assumption of an intrinsic spectrum compatible with a power-law function result in $\tau(E)<1.73\tau_{D11}$, where $\tau_{D11}$ is the optical depth at $z=0.939$ from the Dominguez et al. (2011) model \cite{Dominguez2011}.

\section{Conclusions}

QSO~B0218+357 ($z=0.944$) and PKS~1441+25 ($z=0.939$) were detected for the first time in the VHE range by MAGIC, extending the maximum redshift for VHE $\gamma$-ray observations up to a distance $z\sim 1$, near the cosmic $\gamma$-ray horizon. 
QSO~B0218+357 was the first ever gravitationally lensed blazar detected in the VHE, and the analysis for this peculiar source is still ongoing \cite{Ahnenprep}. The EBL was probed and investigated, obtaining constraints on the existing models for PKS~1441+25. The reduction in the $\gamma$-ray detection threshold operated by MAGIC during the last years thanks to several upgrades and to the improvement of the performance also from the analysis point of view led to the discovery of such distant galaxies and opens the door to the possibility to detect even farther sources at VHE and improve the present knowledge of the $\gamma$-ray Universe.

\subsection{Acknowledgments}
We would like to thank the Instituto de Astrof\'{\i}sica de Canarias
for the excellent working conditions at the Observatorio del Roque de los Muchachos in La Palma.
The financial support of the German BMBF and MPG, the Italian INFN and INAF,
the Swiss National Fund SNF, the ERDF under the Spanish MINECO (FPA2012-39502) and MECD (FPU13/00618), and
the Japanese JSPS and MEXT is gratefully acknowledged.
This work was also supported by the Centro de Excelencia Severo Ochoa SEV-2012-0234, CPAN CSD2007-00042, and MultiDark CSD2009-00064 projects of the Spanish Consolider-Ingenio 2010 programme, by grant 268740 of the Academy of Finland,
by the Croatian Science Foundation (HrZZ) Project 09/176 and the University of Rijeka Project 13.12.1.3.02,
by the DFG Collaborative Research Centers SFB823/C4 and SFB876/C3, and by the Polish MNiSzW grant 745/N-HESS-MAGIC/2010/0.
The {\it Fermi}-LAT Collaboration acknowledges support for LAT development, operation and data analysis from NASA and DOE (United States), CEA/Irfu and IN2P3/CNRS (France), ASI and INFN (Italy), MEXT, KEK, and JAXA (Japan), and the K.A. Wallenberg Foundation, the Swedish Research Council and the National Space Board (Sweden). Science analysis support in the operations phase from INAF (Italy) and CNES (France) is also gratefully acknowledged.
We thank the {\it Swift} team duty scientists and science planners.

\end{document}